\documentclass[aps,prl,floatfix,twocolumn,showpacs,amsmath,amssymb,superscriptaddress]{revtex4-1}
\usepackage{graphicx}
\usepackage{epsfig} 
\usepackage{amssymb}
\usepackage{siunitx}
\usepackage{amsmath}
\usepackage{bm}
\usepackage{color}
\allowdisplaybreaks
\newcommand{\be}{\begin{equation}}
\newcommand{\ee}{\end{equation}}
\newcommand{\bea}{\begin{eqnarray}}
\newcommand{\eea}{\end{eqnarray}}

\def\G{\Gamma}

\def\n{\nu}

\def\bra{\langle}
\def\ket{\rangle}

\begin{document}
\title{Exciton-Peierls mechanism and universal many-body gaps in carbon nanotubes} 
\author{Maria Hellgren}
\affiliation{Sorbonne Universit\'e, Mus\'eum National d'Histoire Naturelle, UMR CNRS 7590, IRD, Institut de Min\'eralogie, de Physique des Mat\'eriaux et de Cosmochimie, IMPMC, 4 place Jussieu, 75005 Paris, France}
\author{Jacopo Baima}
\affiliation{Sorbonne  Universit\'e, UMR CNRS 7588, Institut  des  Nanosciences  de  Paris, INSP, 4 place Jussieu, 75005  Paris,  France}
\author{Anissa Acheche}
\affiliation{Sorbonne Universit\'e, Mus\'eum National d'Histoire Naturelle, UMR CNRS 7590, IRD, Institut de Min\'eralogie, de Physique des Mat\'eriaux et de Cosmochimie, IMPMC, 4 place Jussieu, 75005 Paris, France}
\date{\today} 
\begin{abstract}
"Metallic" carbon nanotubes exhibit quasiparticle gaps when isolated from a screening environment. 
The gap-opening mechanism is expected to be of electronic origin but the precise nature is debated. In this work we show that hybrid density functional theory predicts a set of excitonic instabilities capable of opening gaps of the size found in experiment. The excitonic instabilities are coupled to vibrational modes, and, in particular, the modes associated with the ${\bf \G}-E_{2g}$ and ${\bf K}-A'_1$ Kohn anomalies of graphene, inducing Peierls lattice distortions with a strong electron-phonon coupling. In the larger tubes, the longitudinal optical phonon mode becomes a purely electronic dimerization that is fully symmetry conserving in the zigzag and chiral tubes, but breaks the symmetry in the armchair tubes. The resulting gaps are universal (i.e. independent of chirality) and scale as $1/R$ with tube radius.
\end{abstract}
\maketitle
Low-dimensional carbon systems have been subject of intense research for the last two decades, both for their promise as building blocks in future nanoelectronic devices, and for their novel fundamental physical properties \cite{RevModPhys.88.025005}. 
In reduced dimensions the electron-electron (e-e) interaction is expected to play an increasing role, driving phenomena such as edge magnetism \cite{edgemag}, exciton condensation \cite{macbi} and superconductivity \cite{Tang2462,caotwist}. 
Important examples are the quasi-one-dimensional carbon nanotubes (CNTs) that exhibit a wide range of features depending on their radius and chirality \cite{charlier2007review}. Within a single-particle description, some CNTs are predicted to be semiconducting while others metallic. However, in a pioneering experiment \cite{Deshpande_gap_experiment} it was demonstrated that the e-e interaction induces gaps in all "metallic" CNTs - if suspended in a clean environment. 
This observation was recently confirmed to be universal, i.e., independent of chirality \cite{gaps2018}. 

The physical origin of the observed quasiparticle gaps has not yet been settled. 
First-principle calculations have mainly focused on Peierls transitions, driven by the electron-phonon interaction \cite{kresse-p,blase94,bohnen2004peierls,connet2005peierls,lazzeri-p}. Local functionals in density functional theory (DFT) predict soft phonon modes and lattice instabilities associated with the ${\bf K}-A'_1$ Kohn anomaly of graphene \cite{kohngraph}. The Peierls gaps are, however, an order of magnitude too small and scale as $1/R^3$ with tube radius, in contrast to the approximate $1/R$ behaviour found in experiment. On the other hand, in Ref. \cite{ernzerhof2010peierls} calculations were performed with an improved description of the e-e interaction via hybrid DFT, showing that the picture can change. If the long-range Coulomb interaction is accounted for, small armchair CNTs exhibit Peierls distortions already in the 
unit cell. Although this study points toward larger gaps they again appear to decay quickly with tube radius.

Purely electronic mechanisms such as excitonic instabilities, or electronic charge density waves, also driven by the long-range Coulomb interaction, have been shown to produce the correct scaling with tube radius \cite{effective_low_en,PhysRevLett.79.5086,odintsov,tsvelik,rontani_exc,varsano}.  
Such an instability was recently studied with {\em ab-initio} quantum Monte Carlo (QMC) in the (3,3) armchair tube, and a stable charge-transfer electronic state that breaks the sublattice symmetry was found \cite{varsano}. Calculated gaps were, however, more than an order of magnitude smaller than experimental observations. In addition, most theoretical studies have focused on the armchair family of CNTs. It is, therefore, not clear which theories can account for the weak chirality dependence.

In this work, we present a mechanism based on a unified picture of Peierls distortions and excitonic instabilities. We employ first-principle hybrid DFT that captures the long-range exchange interaction, crucial to describe both excitonic and lattice instabilities in suspended CNTs. We then show that the $A_1$ longitudinal optical phonon mode is unstable, leading to a Peierls distortion, or dimerization, that is fully symmetry conserving in the zigzag and chiral tubes, but breaks the mirror symmetry in the armchair tubes. The lattice distortion is shown to be coupled to a purely electronic instability of the same symmetry, allowing the dimerization to be formed spontaneously already at the electronic level. For the larger tubes the lattice distortion vanishes while the electronic instability remains and becomes degenerate within a multiplet of excitonic states. The corresponding gaps are universal and in good quantitative agreement with experiments.

To the nominally metallic CNTs belong the $(n,n)$ armchair (${\mathcal{A}}$) tubes, the $(3n,0)$ 
zigzag (${\mathcal{Z}}$) tubes, and the $(3l+n,n)$ chiral (${\mathcal{C}}$) tubes ($n$ and $l$ being integers). 
By applying periodic boundary conditions to the band structure of a 
rolled graphene sheet (zone-folding approximation) these tubes inherit 
the two Dirac points formed by the $\pi$ and $\pi^*$ bands of graphene.
The position of the Dirac points depend on the chirality, ranging from 
$K=\pm 1/3$ in ${\mathcal{A}}$-tubes to the two degenerate Dirac points at $\Gamma$ in 
the ${\mathcal{Z}}$-tubes. In the case of ${\mathcal{A}}$-tubes, the Dirac 
point is protected against symmetry-preserving perturbations, while in the ${\mathcal{Z}}$- and ${\mathcal{C}}$-tubes, different mechanisms such as curvature or strain, can open a small gap at the Fermi-level \cite{charlier2007review}. 

An atomic displacement along a phonon mode can also open a gap, provided the electron-phonon coupling (EPC) mixes valence and conduction states \cite{kresse-p,lazzeri-p}. It is well known that in graphene the ${\bf\Gamma}-E_{2g}$ and ${\bf K}-A_1'$ 
phonon modes are significantly softened by the EPC, displaying a Kohn anomaly \cite{kohngraph,wirtzepc}. Within local approximations in DFT, it has been shown that the strong EPC carries over to metallic CNTs and is enhanced by confinement \cite{lazzeri-p,lazzeri-epc}.
The degenerate ${\bf\Gamma}-E_{2g}$ mode splits into a longitudinal optical (LO) and a transverse optical (TO) mode, where the frequency of the LO mode is softened since the distortion along this pattern opens a gap \cite{kresse-p}. 
Similarly, the ${\bf K}-A_1'$ mode of graphene turns into a $2k_F$ mode in the ${\mathcal{A}}$- and ${\mathcal{C}}$-tubes, and a $\G$ mode (which breaks the rotational symmetry) in the ${\mathcal{Z}}$-tubes. Depending on the strength of the EPC, the soft phonon frequency may result in a permanent lattice distortion with a gap, i.e., a Peierls transition. This has been shown to occur for different modes in some small CNTs. 
The strength of the mixing, is, however, sensitive to how the e-e interaction is treated, and, in particular, to the long-range part of the exchange interaction \cite{wirtzepc,tise2}. 
If strong enough, the orbitals could mix spontaneously, i.~e., solely due to the e-e interaction, without the need for an atomic displacement. This would lead to an excitonic insulator state as formulated in Refs. \cite{jerome_excitonic,kozlov1965metal}. 

Let us denote the valence $(\pi)$ and conduction $(\pi^*)$ orbitals of the CNT by $\varphi_{v,k}$ and $\varphi_{c,k}$, where the index $k$ is the 
momentum relative to the valley the orbital belongs to. The mixed orbitals (that must break the symmetry in the $\mathcal{A}$-tubes, but not necessarily in the ${\mathcal{Z}}$-  and ${\mathcal{C}}$-tubes) can be written as linear combinations of the unperturbed orbitals
\bea
\tilde\varphi_{v,k}&=&u_{ k}\varphi_{v,k}+v_{k}\varphi_{c,k}\\
\tilde\varphi_{c,k}&=&u^{*}_{k}\varphi_{c,k}-v^{*}_{ k}\varphi_{v,k}.
\label{orbmix}
\eea 
For a purely electronic instability, the new wavefunction is constructed as an anti-symmetrized product of the 
mixed valence orbitals, i.e.,
\be
|\Psi\ket=\prod_{k}\tilde\varphi^\dagger_{v,k}|0\ket=\prod_{k}(u_k+v_k\varphi^{\dagger}_{c,k}\varphi_{v,k}^{{\color{white}\dagger}})|\Phi\ket
\label{wavefnk}
\ee
where $|0\ket,|\Phi\ket$ are the vacuum, and ``high-symmetry'' or ``single-particle'' ground states, respectively. The last equality is easily derived, and demonstrates the excitonic nature of the new ground state. In a single-particle description the symmetry of the Hamiltonian is always preserved. 
However, within a self-consistent field (SCF) method, such as Hartree-Fock (HF) or DFT, the effective self-consistent HF or 
Kohn-Sham potential may break the symmetry of the 
original Hamiltonian. In this way, a symmetry broken excitonic state can also be described as a single Slater 
determinant built up from the orbitals $\tilde\varphi_{v,k}$, generated by the symmetry broken Hamiltonian. 
However, in standard semi-local approximations within DFT, such as PBE, the long-range exchange interaction, responsible for the electron-hole coupling or excitonic effects, is not well described. By mixing in a fraction $\alpha$ of HF exchange (typically around 25\%), electronic properties generally improve as compared to pure PBE, and excitonic effects are captured \cite{kresse_excitons_hse,ferrari2015}. In the Supplemental Material we demonstrate how an excitonic instability can be captured with a hybrid functional \cite{SuppMat}. 

Although the essential physics is 
described by the $\pi$ and $\pi^*$ bands, in this work we carry out first-principle simulations that include all states of the CNTs. This is achieved with the {\tt CRYSTAL} program \cite{crystal14,noel2009nano_crystal}, that uses
atom-centered Gaussian basis functions, which allow for an efficient evaluation of HF exchange \cite{method,demichelis2011nano_crystal}. Hybrid functionals have been shown to reasonably well reproduce the many-body enhancement of the EPC in graphene \cite{wirtzepc}. In Ref. \cite{ernzerhof2010peierls} 
$\alpha$ was optimized to $30\%$ for CNTs, using the experimentally determined gap and Peierls distortion in {\it trans}-polyacetylene (a 1D, $sp_2$ bonded system with similar electronic structure to CNTs). In this work we, therefore, use this fraction of exchange and denote the approximation by HYB30. 

\begin{figure}[t]
\includegraphics[width=0.99\columnwidth]{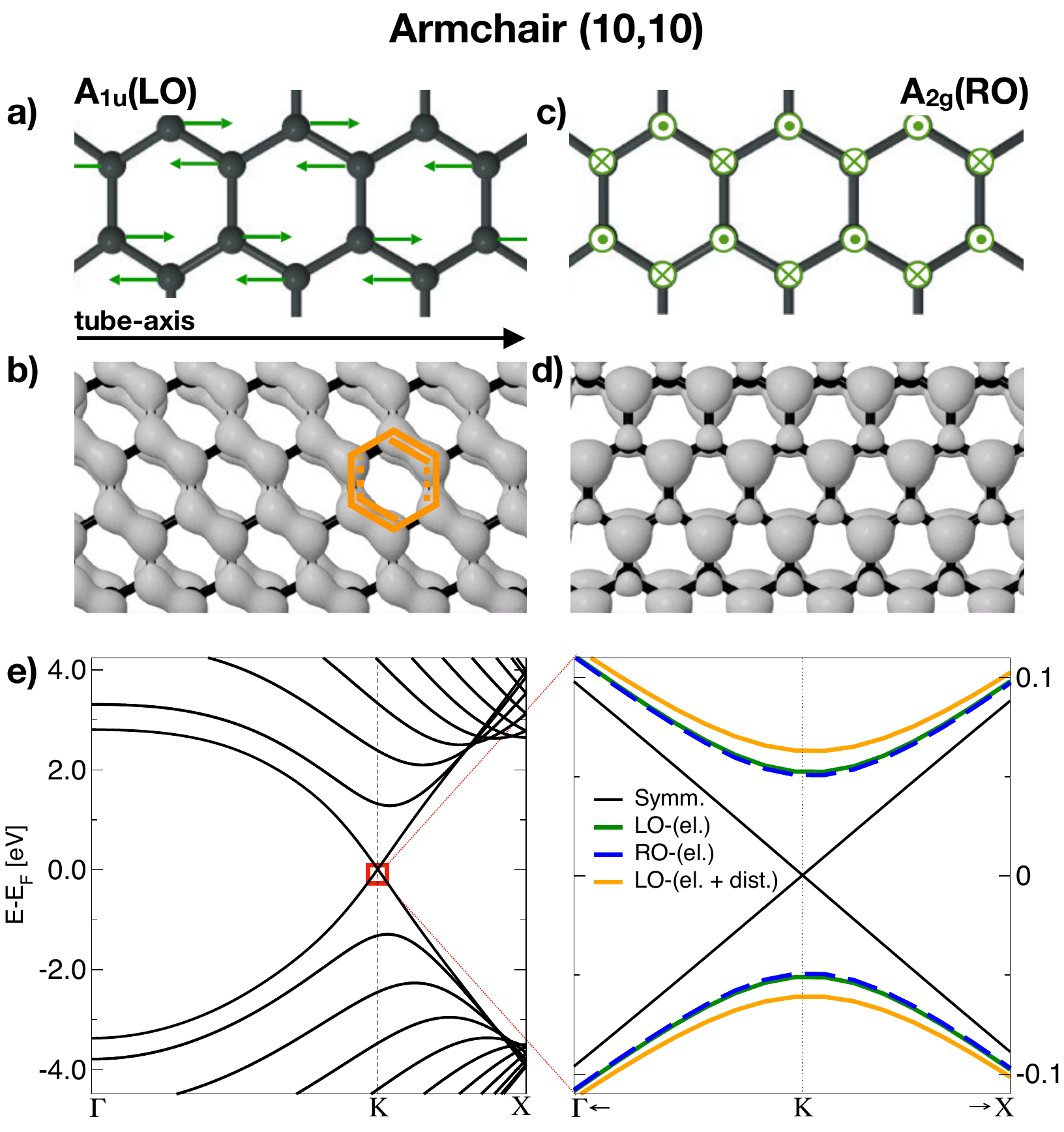}
\caption{(a) and (c) Displacement pattern of the LO and RO modes in the $\mathcal{A}$-tubes. (b) and (d) Density obtained from states within 50 meV below the top of the valence band. 
(e) Left: Band structure of the high-symmetry phase of the (10,10) tube. Right: A zoom around the Fermi-level. The gap of the LO mode is shown with (el. + dist.) and without (el.) lattice relaxation, and compared 
to that of the RO.}
\label{fig:density}
\end{figure}

We start by analyzing the (10,10) $\mathcal{A}$-tube ($R=0.68$ nm), for which the semi-local PBE and the short-range hybrid HSE06 functionals stabilize a gapless symmetric state. With the HYB30 approximation we 
find this state to be unstable with respect to a distortion along the $A_{1u}$(LO) phonon mode (Fig. \ref{fig:density}a). This distortion opens a gap of 124 meV
(orange curve in Fig. \ref{fig:density}e) and breaks horizontal and vertical mirror symmetries. 
 A similar Peierls like distortion was found in the smaller tubes in Ref. \cite{ernzerhof2010peierls},
while within PBE a softened phonon mode is observed \cite{kresse-p}.
If we now fix the lattice at the symmetric configuration (i.e. the undistorted structure) but allow for the same symmetry to break in the SCF potential, we obtain an electronic broken symmetry state, or a bond density wave \cite{tsvelik}, following the same pattern as the Peierls distortion (see Fig. \ref{fig:density}b). 
The dimerized bonding pattern thus occurs spontaneously, prior to the lattice distortion. In the Supplemental Material \cite{SuppMat} we demonstrate that this corresponds to a particular orbital mixing, or an excitonic instability. The purely e-e interaction-induced gap accounts for as much as 
80\% of the gap in the (10,10) tube, as demonstrated in Fig. \ref{fig:density}e. 

A symmetry analysis \cite{symmetry_nanotubes,SuppMat} shows that it is also possible to mix the orbitals with a 
different phase of $v_k$, breaking the sublattice symmetry. 
This results in a charge-transfer between the atoms, as in the "dual" charge density wave state of Ref. \cite{tsvelik}, and corresponds to the solution found by QMC \cite{varsano}. This state is coupled to a radial optical (RO) phonon mode of $A_{2g}$ symmetry, as displayed in Fig. \ref{fig:density}c. As the $A_{1u}$(LO) and $A_{2g}$(RO) mixings differ only by the phase of $v_k$ and descend from a doublet of nearly degenerate excitonic states \cite{SuppMat,darkexciton1,ando}, they are expected to be close in energy. Indeed, starting from the symmetric configuration and by selectively switching off symmetry constraints we could stabilize also the RO state, being only slightly higher in energy. This state is, however, only weakly coupled to the lattice and thus no further stabilization occurs. This can be understood from the fact that the corresponding out-of-plane optical (ZO) mode in graphene has no EPC. The dimerized LO state is therefore found to be the ground state in all $(n,n)$ tubes with $n \geq 10$.  
\begin{figure}[t]
\begin{center}
\includegraphics[width=0.99\columnwidth]{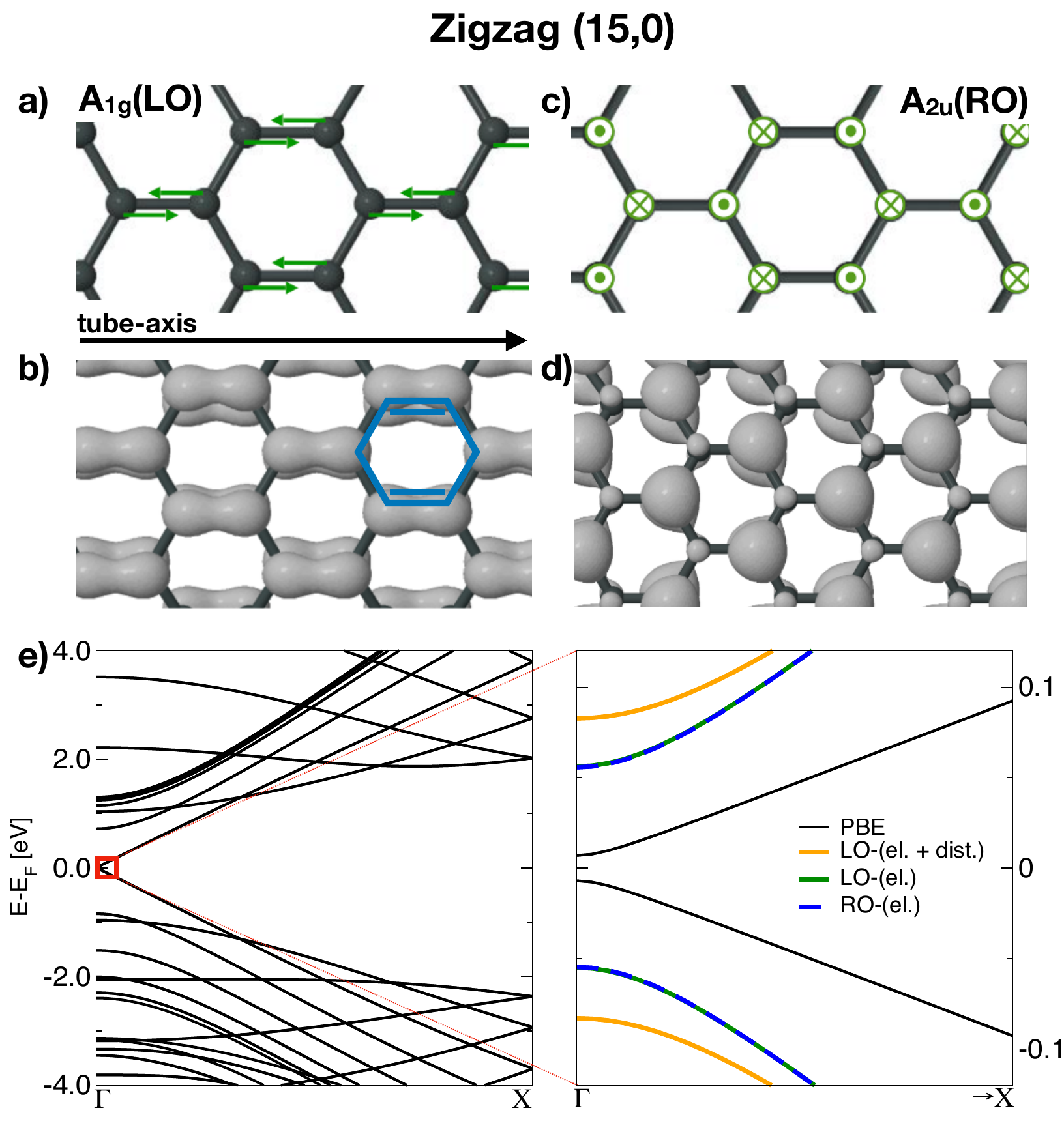}
\caption{(a) and (c) Displacement pattern of the LO and RO modes in the $\mathcal{Z}$-tubes. (b) and (d) Density obtained from states within 50 meV below the top of the valence band. 
(e) Left: PBE band structure of (15,0) tube. Right: A zoom around the Fermi-level. 
The gap of the LO mode is shown with (el. + dist.) and without (el.) lattice relaxation, and compared to that of the RO.}
\label{gap10}
\end{center}
\end{figure}

In the $\mathcal{Z}$-tubes the LO mode is a totally symmetric $A_{1g}$ mode \cite{kresse-p}. In Fig. \ref{gap10} we present the results for the (15,0) tube, which has a similar radius ($R=0.58$ nm) to the (10,10) tube discussed above. In this case the PBE and HSE06 functionals open small gaps of the order of a few meV. The HYB30 functional increases the gap  substantially, up to 165 meV at relaxed geometry. 
It is often assumed that HF theory can only enhance single-particle gaps obtained at the PBE level. In the present case, such an enhancement is, however, expected to be much smaller \cite{aspitarte2017giant}. 
We will now show that the HYB30 gaps are qualitatively different  from the PBE gaps. If we move the atoms along the $A_{1g}$(LO) pattern we find the PBE gaps to vanish, as expected from a single-particle effect.
In contrast, if we use HYB30 only a minimum is reached, at around 110 meV (corresponding to 65\% of the gap at relaxed geometry). To the left in Fig. 2b we show the dimerized charge density close to the Fermi energy. This is the $A_{1g}$(LO) dimerized pattern.
Despite the absence of symmetry breaking, the physical mechanism is the same as in the $\mathcal{A}$-tubes, i.e., a dimerization along the LO mode. 
The similarity to $\mathcal{A}$-tubes is further emphasized by the fact that we also find a charge transfer solution at higher energy. In order to induce this state we have to remove the preference determined by the EPC and start from the minimum-gap geometry with the proper symmetry constraints switched off (see Fig. 2d). 

The $\mathcal{C}$-tubes have larger unit cells and less symmetry than the $\mathcal{A}$ -and $\mathcal{Z}$-tubes, and exhibit only near LO and TO modes \cite{lazzeri-p}. The near LO mode is a totally symmetric mode like in the $\mathcal{Z}$-tubes. The $\mathcal{C}$- and $\mathcal{Z}$-tubes, thus, behave very similarly. In this work we have called the metal-insulator transition (MIT) excitonic due to the importance of long-range interactions, but we note that the excitonic insulator 
state is usually assumed to involve symmetry-breaking \cite{kohn_excitonic}. Symmetry conserving MITs of electronic origin are often called Mott transitions.

\begin{figure}[t]
\begin{center}
\includegraphics[width=\columnwidth]{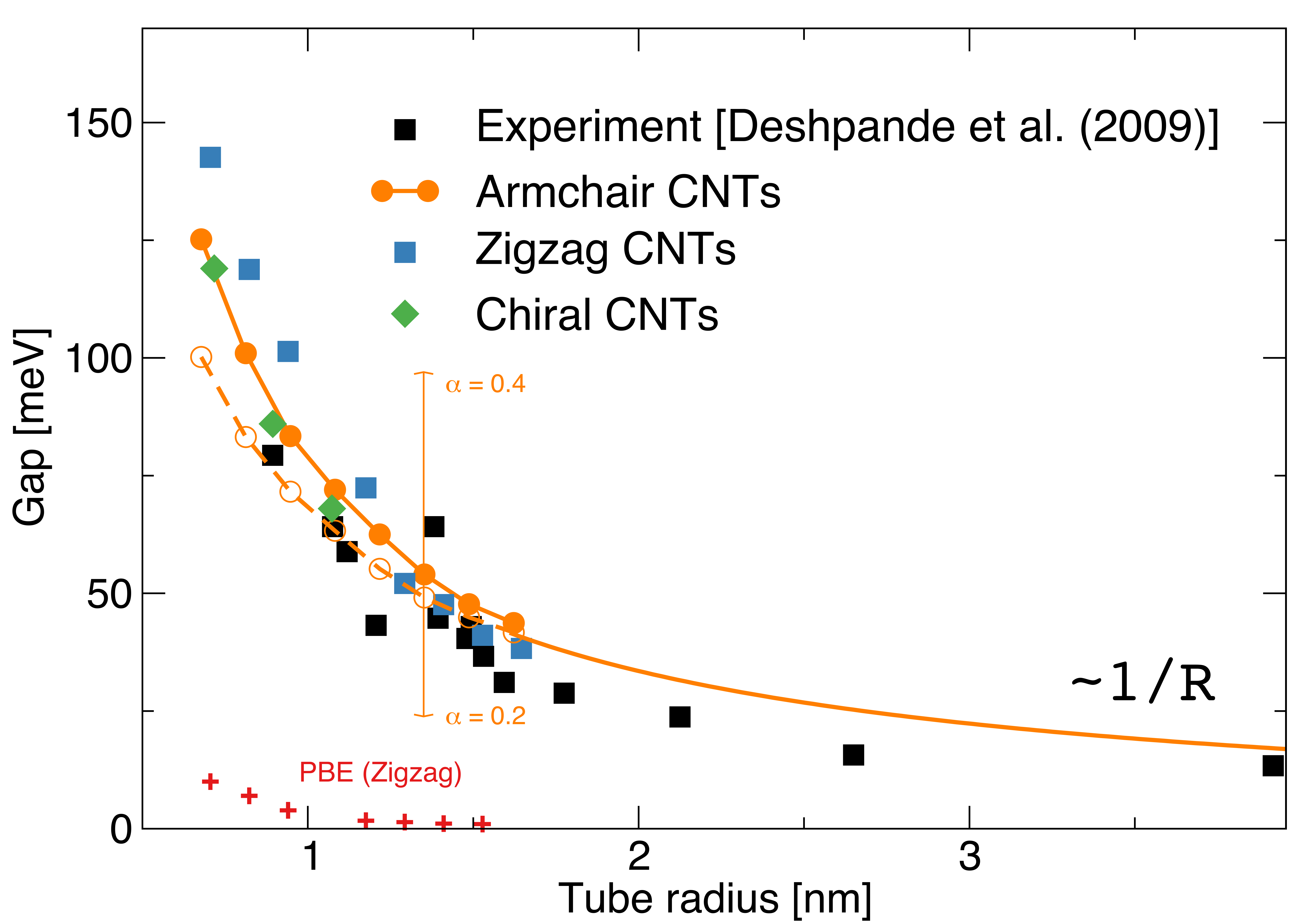}
\caption{Exciton-Peierls gaps in comparison with experiment (black squares). The orange circles (blue squares) correspond to the fully relaxed states of the $\mathcal{A}$($\mathcal{Z}$)-tubes. The dashed orange line is an interpolation of the electronic gaps of the $\mathcal{A}$-tubes, extrapolated with a $1/R$ fit (full orange line). The green diamonds are $\mathcal{C}$-tubes [(16,4),(20,5),(24,6)].}
\label{gaps}
\end{center}
\end{figure}
We now extend the analysis to larger CNTs ($R > 0.7$ nm). In Fig. \ref{gaps} we compare our computed gaps for the fully relaxed LO states in the $\mathcal{A}$-, $\mathcal{Z}$- and some $\mathcal{C}$-tubes, with the low-temperature experiment performed by Deshpande et al. \cite{Deshpande_gap_experiment}. 
Our results match experiment rather well and in this range of radii the calculated gaps are close to independent of chirality. The presence of a lattice instability induces a small deviation from the excitonic $1/R$ behaviour \cite{varsano} at low enough temperature. Around $R = 1.5$ nm, the lattice contribution vanishes and the gaps become purely electronic. The electronic gaps of the $\mathcal{A}$-tubes (dashed orange line) follow a perfect $1/R$ behavior in the whole range, which allows us to extrapolate the results for $R>1.6$ nm, indicated by the full orange line. The electronic states in the larger CNTs are highly degenerate, possibly infinitely degenerate (i.e. a complete phase invariance of the ground state). We note, however, that a strong EPC is likely to destroy super-transport properties associated with the excitonic ground state \cite{fate_excitonic}. We also recall the approximate nature of the hybrid functional which tend to underestimate the lattice distortion \cite{ernzerhof2010peierls}. 
The magnitude of the gaps varies as $\alpha^2$. To illustrate how a variation of $\alpha$ affects the gaps we have added an 'error-bar' in Fig. \ref{gaps}. Within a realistic range ($0.2 < \alpha < 0.4$), the gaps are compatible with all experimental results presented so far \cite{Deshpande_gap_experiment,gaps2018}. We note that even on the lower bound we have a substantial gap-opening.  
\begin{figure}[t]
\begin{center}
\includegraphics[width=\columnwidth]{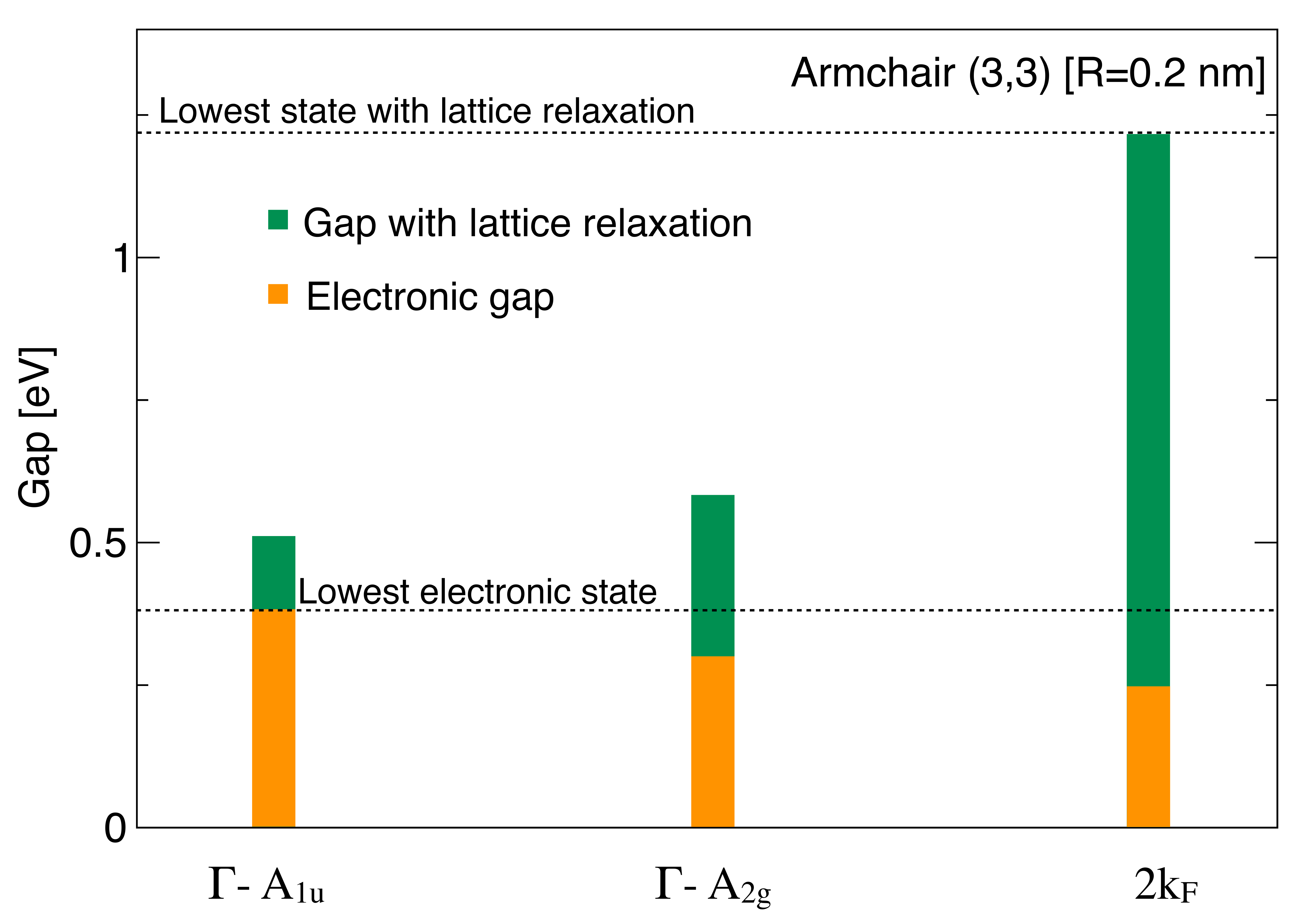}
\caption{Comparison between gaps from the different instabilities in the (3,3) tube. 
The gaps are decomposed into the electronic part (orange) and the part obtained after lattice relaxation (green). In all cases larger gaps correspond to more stable states.}
\label{nano3}
\end{center}
\end{figure}

We now turn to the  smaller CNTs for which lattice instabilities become increasingly important. The electronic states split in energy already by the e-e interaction, with the $A_1$-Peierls state always lower in energy.
The splitting becomes larger the smaller is the tube, 
similarly to what is found for excitonic states in small semi-conducting CNTs \cite{darkexciton1,ando,matsunaga2008KKprime,molinari_dark_excitons}. For the small nanotubes we also find a structural relaxation along the $A_2$(RO) phonon mode associated with the charge-transfer electronic state. We focus on the (3,3) tube, which is interesting in its own right as CNTs of this radius have been reported to superconduct up to 15 K \cite{Tang2462}. In addition to the $\G$-transitions we also study the $2k_F$-instabilities. As $2k_F$ is incommensurate with the lattice we can, however, only 
study this instability approximately. We found that for the (3,3) tube a cell which is 7 times 
larger than the original cell approximates $2k_F$ within $1.1\%$. By relaxing the density 
in this supercell we found, indeed, a symmetry broken solution of electronic origin \cite{SuppMat}. 
At the electronic level this state is less stable than the ones at $\G$, as shown in Fig. \ref{nano3}, which again follows the trend of the splitting between exciton binding energies \cite{ando}.  However, the strength of the electron-lattice interaction in such a small tube is 
very large and a lattice instability is present even with local functionals \cite{connet2005peierls,bohnen2004peierls}. 
As can be seen in Fig. \ref{nano3}, the interaction with the lattice increases the $2k_F$ gap by a factor of 5 and inverts the stability order, turning the $2k_F$ instability into the lowest state for the (3,3) tube. The role of the long-range e-e interaction is, however, 
still very important. In Tab. S1 of the Supplemental Material \cite{SuppMat} we compare our results for the phonon frequencies to 
PBE data from Ref. \cite{bohnen2004peierls}. We see that the instability with HYB30 is substantially stronger and broader in reciprocal space. With increasing radius 
the $2k_F$ lattice contribution reduces, however, quickly, turning also the $2k_F$ instability purely electronic and degenerate with the $\G$-instabilities.

In conclusion, we have formulated a universal gap-opening mechanism in metallic CNTs based on dimerization that takes into account both electronic and lattice degrees of freedom. In the $\mathcal{Z}$ -and $\mathcal{C}$-tubes the gaps appear without any symmetry breaking, while in the $\mathcal{A}$-tubes they are induced by a Peierls lattice instability that breaks the symmetry. We have shown that these seemingly different phenomena share the same origin in an electronic transition associated with the $A_1$(LO) phonon mode.
While the electron-lattice interaction is crucial in the smaller tubes, in the experimentally observed range of diameters the electronic transition dominates and yields gaps that can reproduce experimental findings.
\begin{acknowledgments}
This work was performed using HPC resources from GENCI-TGCC/CINES/IDRIS (Grant 2017- DARI A0010907625).
 \end{acknowledgments}
%
\cleardoublepage 

\begin{widetext}
\section{Supplemental material for: \\
Exciton-Peierls mechanism and universal many-body gaps in carbon nanotubes}
Below we present further details on the Hartree-Fock (HF) excitonic ground state, the symmetry of 
the CNT Bloch orbitals, and supplementary data on the $2k_F$ instability in the (3,3) tube.

\section{Excitonic instabilities within Hartree-Fock theory}  
The wave function of the excitonic ground-state is defined as a Slater determinant built up from 
the symmetry broken orbitals. These orbitals can be expressed as linear combinations of 
valence (v) and conduction (c) orbitals from a symmetry restricted calculation. Orbitals can 
be mixed with the same momentum, breaking a point group symmetry, or they can differ by momentum $2k_F$, 
breaking the translational symmetry. We can write the new valence orbitals as
\bea
\tilde\varphi_{v,k}&=&u_{ k}\varphi_{v,k}+v_{k}\varphi_{c,k},
\label{orbital}
\eea 
where $k$ is the momentum relative to the valleys $K$ or $K'$ and spin-indices have been suppressed 
(only spin-singlet states are considered). 
We then form a new Slater determinant 
\be
|\Psi\ket=\prod_{k}\tilde\varphi^\dagger_{v,k}|0\ket.
\ee
To determine whether this new ansatz lowers the energy one should minimize the expectation value 
$\bra\Psi |\hat H|\Psi\ket$ with respect to the coefficients $v_{k},u_{k}$. $\hat H$ is 
the many-body Hamiltonian given by
\be
\hat H = \hat T + \hat V_{\rm ext} + \hat W
\label{hamei}
\ee
where $\hat T$ is the kinetic energy operator, $ \hat V_{\rm ext}$ is the external lattice potential and $\hat W$ is the interaction 
operator of the electron-electron interaction ($v$). The expectation value of $\hat W$ is the standard HF result
\be
\bra\Psi |\hat W|\Psi\ket=\frac{1}{2}\sum_{kk'}\bra \tilde\varphi_{v,k}\tilde\varphi_{v,k'}|v|\tilde\varphi_{v,k}\tilde\varphi_{v,k'}\ket-\bra \tilde\varphi_{v,k}\tilde\varphi_{v,k'}|v|\tilde\varphi_{v,k'}\tilde\varphi_{v,k}\ket
\ee
Assuming that the orbitals $\varphi_{v,k},\varphi_{c,k}$ are solutions to the HF equations of a symmetry restricted calculation we can write 
\bea
\bra\Psi |\hat H|\Psi\ket & =&\sum_{k}|u_{k}|^2\varepsilon_{v,k}+\sum_{k}|v_{k}|^2\varepsilon_{c,k}\\
&& +\frac{1}{2}\sum_{kk'} |u_k|^2[|u_{k'}|^2-2][\bra \varphi_{v,k}\varphi_{v,k'}|v|\varphi_{v,k}\varphi_{v,k'}\ket-\bra \varphi_{v,k}\varphi_{v,k'}|v|\varphi_{v,k'}\varphi_{v,k}\ket ]
\\
&& + \frac{1}{2}\sum_{kk'} |v_k|^2|v_{k'}|^2[\bra \varphi_{c,k}\varphi_{c,k'}|v|\varphi_{c,k}\varphi_{c,k'}\ket-\bra \varphi_{c,k}\varphi_{c,k'}|v|\varphi_{c,k'}\varphi_{c,k}\ket ]
\\
&& + \sum_{kk'} |v_{k}|^2[|u_{k'}|^2-1][\bra \varphi_{v,k'}\varphi_{c,k}|v|\varphi_{v,k'}\varphi_{c,k}\ket-\bra \varphi_{v,k'}\varphi_{c,k}|v|\varphi_{c,k}\varphi_{v,k'}\ket ]
\\
&& + \frac{1}{2}\sum_{kk'}  u_k^*u_{k'}^*v_kv_{k'}[\bra \varphi_{v,k}\varphi_{v,k'}|v|\varphi_{c,k}\varphi_{c,k'}\ket-\bra \varphi_{v,k}\varphi_{v,k'}|v|\varphi_{c,k'}\varphi_{c,k}\ket ]+{\rm c.c.}
\\
&& +\frac{1}{2}\sum_{kk'}  u_k^*v_{k'}^*v_ku_{k'}[\bra \varphi_{v,k}\varphi_{c,k'}|v|\varphi_{c,k}\varphi_{v,k'}\ket-\bra \varphi_{v,k}\varphi_{c,k'}|v|\varphi_{v,k'}\varphi_{c,k}\ket ]+{\rm c.c.}
\\
&& + \sum_{kk'} u_k^*|v_{k'}|^2v_k[\bra \varphi_{v,k}\varphi_{c,k'}|v|\varphi_{c,k}\varphi_{c,k'}\ket-\bra \varphi_{v,k}\varphi_{c,k'}|v|\varphi_{c,k'}\varphi_{c,k}\ket ]+{\rm c.c.}
\\
& &+\sum_{kk'}  v_k^*u_k[|u_{k'}|^2-1][\bra \varphi_{c,k}\varphi_{v,k'}|v|\varphi_{v,k}\varphi_{v,k'}\ket-\bra \varphi_{c,k}\varphi_{v,k'}|v|\varphi_{v,k'}\varphi_{v,k}\ket ]+{\rm c.c.}
\eea
A variation with respect to the expansion coefficients $u_k,v_k$ yields the HF equations of the excitonic ground-state.\cite{jerome_excitonic} 
We see that the same expressions hold in the case of a hybrid functional by simply multiplying the Fock term with the fraction of 
exchange and by adding a single-particle PBE term. In model Hamiltonians\cite{jerome_excitonic,fate_excitonic} only terms which are 
invariant with respect to a $k$-independent phase of $v_k$ are kept (terms in Eqs. (7-10,12) above), but we see here that, in general, 
there are terms which are not invariant with respect to this phase. Indeed, for small CNTs we find a splitting at the electronic level 
between the $A_{1u}$(LO) and the $A_{2g}$(RO) states (two states that differ by a phase angle of $\pi/2$, see below). This splitting gets smaller with 
the size of the tube and eventually vanishes, indicating that for the larger tubes $R>1.6 {\rm \ nm}$ the phase dependent terms vanish.
\section{Excitonic broken-symmetry states in $\mathcal{A}$-tubes}
The symmetry group of $(n,n)$ armchair nanotubes includes the following
symmetry operations:\cite{symmetry_nanotubes} 
\begin{itemize}
\item an $n$-fold rotation around the axis of the tube
\item infinite rototranslational symmetry, generated by a translation of $a/2$ combined with a rotation of $\pi/n$ around the tube axis, which generates the periodicity
\item a vertical and an horizontal mirror, $\sigma_v$ and $\sigma_h$ crossing at the center of each hexagon, plus a 180 degrees rotation 
$\textup{U}=\sigma_v  \sigma_h$. These symmetry operations are represented in Fig. \ref{fig:symmetry}.
\end{itemize}

\begin{figure}[ht]
\includegraphics[width=0.4\columnwidth]{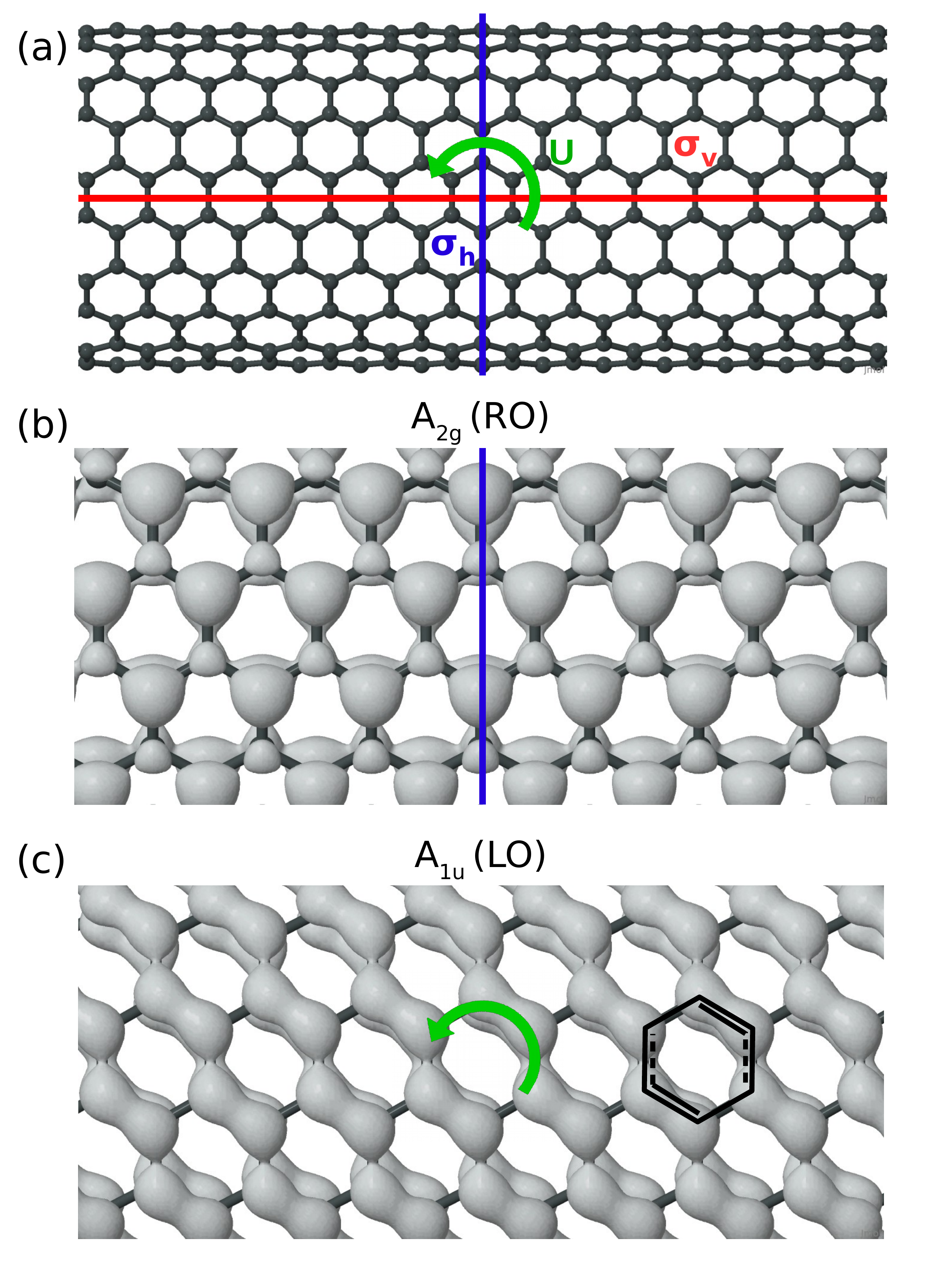}
\caption{Panel (a) shows the lattice of the (10,10) nanotube with symmetry operators indicated. Panel (b)
and (c) show the density of the $A_{1u}$ and $A_{2g}$ purely electronic broken 
symmetry states. The symmetry that is preserved is indicated. }
\label{fig:symmetry}
\end{figure}

It should be noted that applying either operator $\sigma_h$ or U inverts the
periodic direction.
The atoms composing the nanotube can be divided into two
sublattices A and B, with the first neighbours of each atom always belonging to
a different sublattice. Applying either operator $\sigma_v$ or U exchanges the
two sublattices.

Let us consider the crystalline orbitals of the symmetry-restricted problem
in the highest valence band ($\pi$) $\varphi_{v,k}$ and the lowest 
conduction band ($\pi^*$) $\varphi_{c,k}$. These are in the same representation of the 
rotational subgroup and have opposite parity $\Pi=\pm 1$ with respect
to $\sigma_v$ (the valence band 
$\varphi_{v,k}$ has parity $\Pi=-1$ from $\Gamma$ to K and $\Pi=+1$ 
from K to border zone, while the inverse is true for $\varphi_{c,k}$).\cite{symmetry_nanotubes}
The wavefunction for a $\G$ exciton can be expressed as:
\be 
\Psi_{exc}=\sum_k A_k \varphi_{v,k}^* \varphi_{c,k}. 
\ee
Due to the symmetry of the orbitals, $\Psi_{exc}$ is symmetric with respect
to rotations and antisymmetric with respect to $\sigma_v$. The relationship $\textup{U}=\sigma_v \sigma_h$ implies that the
exciton wavefunction has opposite parity with respect to $\textup{U}$ and $\sigma_h$.
Moreover, the effect of symmetry operators U and $\sigma_h$
on the orbitals can be written as:\cite{symmetry_nanotubes}
\be
\sigma_h\varphi_{i,k}=\varphi_{i,-k} \text{ \; \; } , \text{ \; \; } U\varphi_{i,k}=\Pi \varphi_{i,-k}
\ee
with $i=v,c$. It is easy to verify that the parity of $\Psi_{exc}$
with respect to U is even if $A_k=-A_{-k}$ (i.e. if the exciton wavefunction is antisymmetric with respect to exchanging the two valleys at K and -K) and odd if $A_k=A_{-k}$, as in the case of chiral semiconducting nanotubes.\cite{ando,matsunaga2008KKprime} These are
often referred to as gerade and ungerade excitons. In semiconducting tubes, these excitons are slightly split in energy, and form a multiplet toghether with two excitons at $2K$.
Note that, if the two excitons are exactly degenerate, they generate a subspace of allowed
excitations with no definite parity with respect to $\textup{U}$ and $\sigma_h$. 

Similar symmetry considerations apply to the valence states 
$\tilde\varphi_{v,k}$ in Eq. (\ref{orbital}), which are obtained by mixing valence 
and conduction orbitals. If the e-e interaction is responsible for the mixing, these 
states can be interpreted as a spontaneous condensation of 
excitons, as mentioned in the main text.
The charge density of the $\tilde\varphi_{v,k}$ mixed state is:
\be
n(\mathbf{r})=\sum_k|\tilde\varphi_{v,k}(\mathbf{r})|^2=\sum_k |u_k|^2 |\varphi_{v,k}(\mathbf{r})|^2 + \sum_k|v_k|^2 |\varphi_{c,k}(\mathbf{r})|^2 + 2 \sum_k {\rm Re}\left[u^*_k v_k \varphi_{v,k}^*(\mathbf{r})\varphi_{c,k}(\mathbf{r})\right].
\label{density}
\ee
The first and second terms are symmetric with respect to the full symmetry group. The third term resembles the exciton wavefunction,
and the same symmetry considerations apply to it. In particular, it is
asymmetric with respect to $\sigma_v$ and the symmetry properties with
respect to $\sigma_h$ and U depend on the relationship between $v_k$ and $v_{-k}$, assuming that $u_k$ is real and positive 
(equivalent to fixing the overall phase of the wavefunction).

The relative phase between $v_k$ and $v_{-k}$ is then restricted only by time-reversal symmetry:
\be
\tilde\varphi_{v,-k}=\tilde\varphi_{v,k}^*=
u^*_{ k}\varphi_{v,k}^*+v_{k}^*\varphi_{c,k}^*=u^*_{ k}\varphi_{v,-k}+v_{k}^*\varphi_{c,-k}
\implies v_{-k}=v_{k}^*
\ee
Thus if $v_{k}$ is purely real, $v_{-k}=v_{k}$ and hence the third term of Eq. (\ref{density}) 
is antisymmetric with respect to U and symmetric with respect to $\sigma_h$ (as in the case of the ungerade exciton). 
If $v_{k}$ instead is purely imaginary (i.e., having a phase $e^{\pm i\pi/2}$), $v_{-k}=-v_{k}$ the term has the same symmetries 
as the gerade exciton.

\begin{figure}[ht]
\includegraphics[width=0.4\columnwidth]{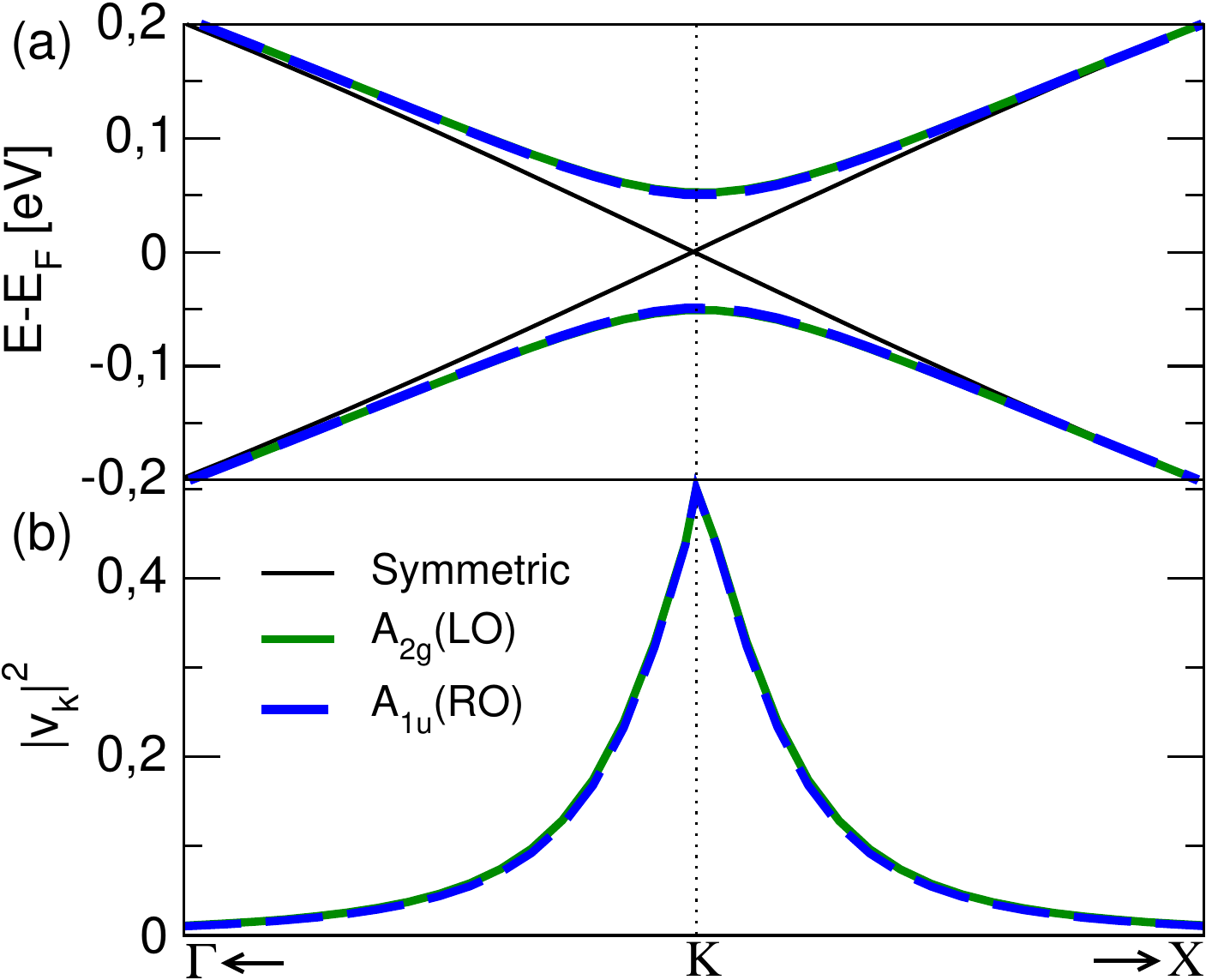}
\caption{(a) A zoom around the Fermi-level of the band structure of the (10,10) $\mathcal{A}$-tube, comparing the symmetric solution with the $A_{2g}$ charge transfer solution (blue) and the  $A_{1u}$-LO one (green). In panel (b), the orbital hybridisation, 
i.e. the mixing between valence and conduction states of the high-symmetry phase is shown. }
\label{fig:mixing}
\end{figure}

The total charge density $n(\mathbf{r})$ has no definite symmetry with respect to $\sigma_v$ whenever $v_k$ is finite, 
since it is the sum of two symmetric and one antisymmetric term.
If $v_{k}$ is real, the symmetry U is also broken allowing a charge-transfer
between the two sublattices (the $A_{2g}$(RO) solution in the main text). If $v_{k}$ is imaginary, $\sigma_h$ is instead
broken but U is preserved and the sublattices are symmetry equivalent. Indeed,
in this case all atoms in the CNT are still related to each other by
symmetry operations, so no charge-transfer between atoms is possible.
However, this state breaks the symmetry between the bonds connecting
each atom to its neighbours along the periodic direction, resulting in a dimerization of the electronic ground state (the $A_{1u}$(LO) solution in the main text). 

In Fig. \ref{fig:mixing} we present the calculated orbital hybridization, i.e., the degree of mixing between the valence and conduction bands (Eq. (1) in the main text) given by
\be
|v_k|^2=|\bra \varphi_{c,k} |\tilde\varphi_{v,k}\ket |^2.
\ee
The $A_{2g}$ and $A_{1u}$ electronic symmetry broken states have almost identical mixing and can thus only differ by a $k$-dependent phase factor $v_k\rightarrow e^{i\theta_k}v_k$, in agreement with the symmetry analysis presented above.
 
\section{Supplementary data on the $2k_F$ instability}
In Tab. \ref{2kf} we have quantified the difference between the strenght of the $2k_F$ Peierls distorsion in the (3,3) tube between 
a local functional (PBE) and the 30\% hybrid that includes the long-range e-e interaction. 
Starting from the relaxed $\Gamma$ geometry (the RO state for this tube) we calculate 
the hybrid phonon frequencies for the unstable mode at different values of $k$, achieved 
by using supercells of different sizes. The results are compared to 
PBE data from Ref. \onlinecite{bohnen2004peierls} taken at the same distance from $2k_F$. 
We see that the instability with the hybrid functional is substantially stronger and 
broader in reciprocal space, in addition to the fact that it is preceded by an 
electronic transition.

\begin{figure}[t]
\includegraphics[width=0.6\columnwidth]{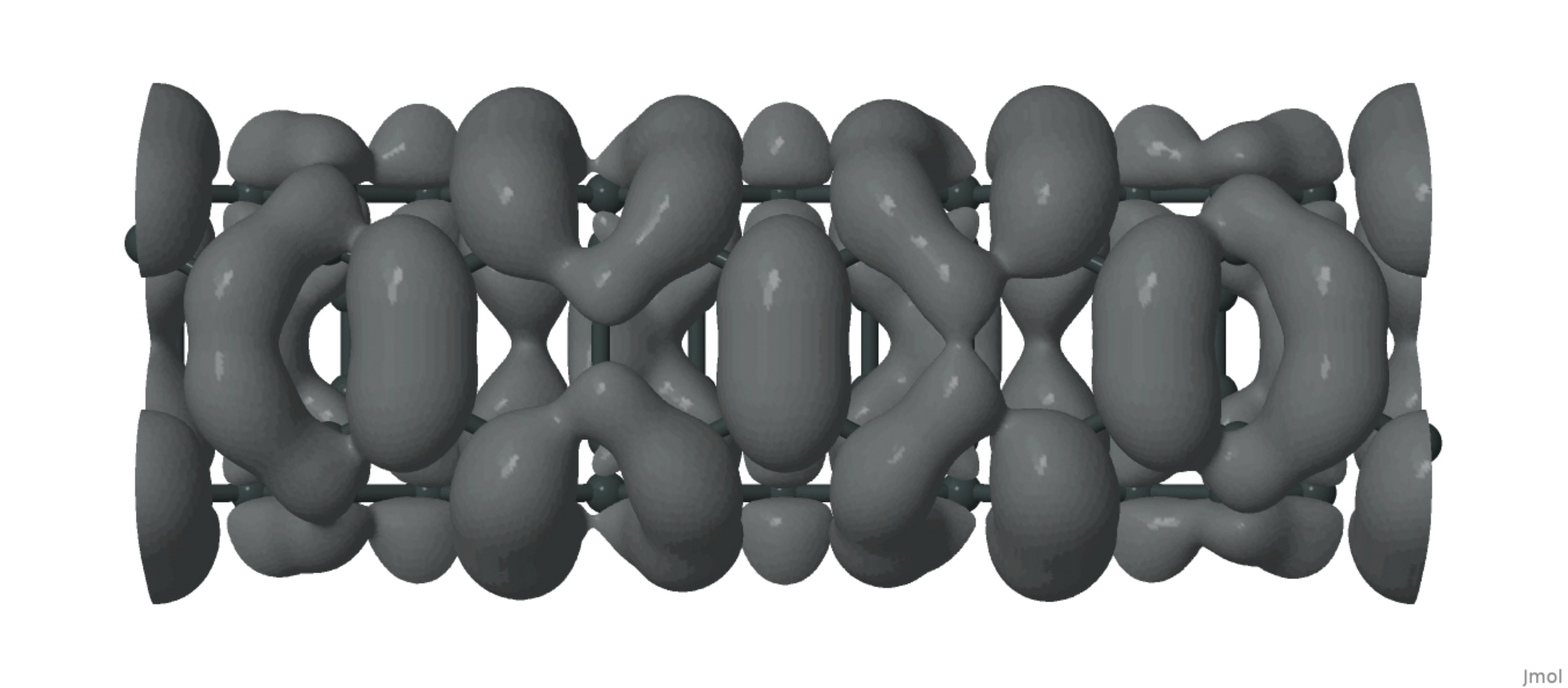}
\caption{Isosurface of the density obtained from the states close to the Fermi level in the $2k_F$ excitonic ground state
of the (3,3) nanotube, having the same symmetry as the $2k_F$ Peierls instability.}
\label{fig:density2kf}
\end{figure}
\begin{table}[t]
\caption{$2k_F$ instability of the (3,3) nanotube: comparison of the phonon 
frequencies obtained with PBE and the hybrid with 30\% of exchange (HYB30). 
The frequencies are computed at the relaxed cell-periodic geometry. The supercells (SC) approximate 
$2k_F$ at the same distance from $2k_F$ as the PBE results extracted from Bohnen {\it  et al.}. The wavevectors $k$ are 
expressed in units of $2\pi/a$.  }\label{2kf}
\begin{ruledtabular}
\begin{tabular}{l ccccccccccc }
$k-2k_F$ &  SC    & $k_{\textup{HYB30}}$  
& $\n_{\textup{HYB30}}$(cm$^{-1}$) & $k_{\textup{PBE}}$ & $\n_{\textup{PBE}}$(cm$^{-1}$)\cite{bohnen2004peierls}.   \\
  \hline
-0.017 &  5 & 0.400   & $824 i$ & 0.412  & 371 \\
0.011 &  7 & 0.429  & $1256 i$ & 0.434  & 282 \\
0.027 &  9 & 0.444   & $545 i$ & 0.456  & 569 \\  
  \end{tabular}
\end{ruledtabular}
\end {table}
\end{widetext}
\end{document}